# Revealing two heat-annealing related photoproduct systems and widely existed subgrain domains in organolead perovskite

*Wei Wang,* [†,§] *Yu Li,* [†,§] *Xiangyuan Wang,* [‡] *Yang Liu, Yanping Lv,* [‡] *Shufeng Wang,\**[,†] *Kai Wang,* [‡] *Yantao Shi,\**[,‡] *Lixin Xiao,* [†] *Zhijian Chen,* [†] *and Qihuang Gong\**[,†] and Collaborative Innovation Center of Extreme Optics, Shanxi University, Taiyuan, Shanxi 030006, China.

[†]State Key Laboratory for Artificial Microstructure and Mesoscopic Physics, Department of Physics, Peking University, Beijing 100871, China.

[‡]State Key Laboratory of Fine Chemicals, School of Chemistry, Dalian University of Technology, Dalian, Liaoning 116024, China.

AUTHOR INFORMATION

[†]Wei Wang and Yu Li contribute equally to this work.

**Corresponding Author**

\* wangsf@pku.edu.cn; shiyantao@dlut.edu.cn; qhgong@pku.edu.cn

## ABSTRACT

For highly interested organolead perovskite based solar cells, the photoproducts are regarded as the co-existed exciton and free carriers. In this study, we carefully re-examined this conclusion with our recently developed density-resolved spectroscopic method. Heat-annealing related two photoproduct systems are observed. We found that the widely accepted model is only true for single crystal and freshly made films without heat-annealing. For those sufficiently heat-annealed films, another system presenting significant emissive exciton-carrier collision (ECC) is discovered. In addition, the appearing of ECC indicates the emerging of an internal morphology after heat annealing, which is assigned to a recently discussed twinning subgrain structure. We proved that such subgrain structures broadly exist in perovskite films. This finding could prove the morphological basis for high performance of perovskite working layers.

## TOC GRAPHICS

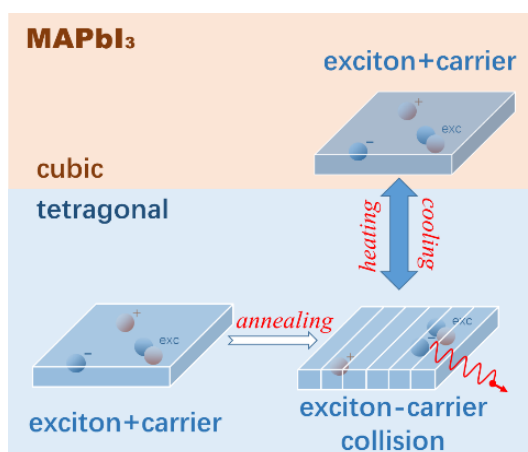

The organolead perovskites are currently the most promising photovoltaic materials. The top efficiency of perovskite based solar cells is beyond 22%.[1] It benefits from rich free carriers and their long diffusion distance up to microns.[2-4] It has been proposed and proved that the exciton and free carriers dynamically co-exist within the perovskite films.[5-6] However, this model seems incomplete because that the extremely low recombination rate of free electron and hole is a general property for the organolead perovskite working layers. A study presented a relative high barrier of 75meV for electron-hole recombination,[7] indicating that the carriers are not moving or recombining freely inside the material. Theoretical analysis suggested that the electron and hole are spatially separated and transported in different tunnels made by ferroelectric domain walls.[8] These discussions indicated that the real situation should be much more complicated than a simple material with regular lattice. A more reasonable analysis should include internal domains at the scale between the crystal lattice and crystal grains. In fact, the subgrain domains had been widely investigated in inorganic perovskites.[9-11] Very recently, ferroelastic/ferroelectric subgrain domains were discussed in organolead perovskite.[12-13] These subgrain structure should present substantial influence to the photophysical behavior as discussed. However, it had not been found yet. In addition, its wide existence is hard to be proved.[13]

It is already known in material science that the photoproduct and morphology are correlated. E.g., in low dimensional semiconductors, some intrinsic hidden photoproducts become visible, such as in $CsPbX_3$ perovskite quantum dot system, where the charged-exciton and bi-exciton were found.[14] The relationship between morphology and photoproduct system is cause and effect, respectively. Though it is possible that the change of morphology may not lead to the variation on photoproducts, the change of the photoproducts must come from the variation on morphology. Therefore, searching the variation on photoproduct system is a reasonable way to disclose the change of morphology.

In this report, we examined the photoproducts in differently processed perovskite films, with our recently developed density-resolved spectroscopic method.[6] This method monitors the density-dependent behavior of photoproducts to directly identify them and their interconversion. In the freshly made, non-heat annealed $CH_3NH_3PbI_3$ film, exciton and free carriers interconvert to each other and formed a dynamically balanced photoproduct system (System I). This is the normally accepted model. However, when the films are sufficiently heat annealed, the free carrier and exciton do not co-exist peacefully. An additional new decay channel of emissive exciton-carrier collision (ECC) appears and become dominant at high excitation density (System II). The analysis and experimental proofs show that a subgrain domain morphology controls the emerging of System II. The morphology is attributed to a ferroelastic/ferroelectric domain structure discussed very recently.[12-13, 15] The universality and robustness of the System II proves that the subgrain domain widely exists. Then the morphology basis for universal high efficient carrier transportation is substantially proved. Both our new experimental method, the newly found photoproduct system, and the proved wide existence of subgrain structure are crucial for investigating the perovskite photophysics and device performance on a subgrain morphological basis.

The density-resolved spectroscopic method is described as following. The $CH_3NH_3PbI_3$ perovskite films were excited at 517 nm. The excitation densities, $n$, were calculated according to the injected photons per pulse and the illuminated volume of the thin films. Due to the fast and long range hot-carrier transportation, the photoproducts are supposed to be uniformly distributed in the volume.[16] The $n$ is between $1\times10^{16}$ and $5\times10^{18}$ $cm^{-3}$. As shown in Fig. 1, the spectra at temporal photoluminescent maxima, $PL_0$, is collected. A detailed description can be found in our former report.[6]

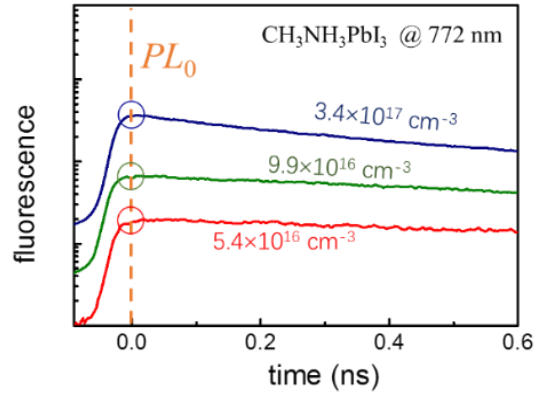

Figure 1. The typical fluorescent decay of $CH_3NH_3PbI_3$ films after excitation. The films were pumped by a femtosecond laser at 517 nm with various pulse energy. The $PL_0$ is taken at the maxima for each fluorescent decay (circled). The corresponding excitation densities are labelled for each decay.

The samples are $CH_3NH_3PbI_3$ (I3) perovskite films, which were either non-heat annealed (I3-na) or with sufficient annealing (I3-sa). A piece of single crystal was also tested. The amplitude of $PL_0$ has power law dependency to $n$, as shown in Fig. 2. When $n$ is lower than $10^{17}$ cm$^{-3}$, the dependencies are quadratic for every sample. At higher density, the dependencies are clearly different. The I3-na, the single crystal, and I3-sa measured at 340K have a linear dependency (Fig. 2 (a), (c), and (d) respectively), while the I3-sa measured at room temperature presents a 3/2 power index towards $n$. Similar power index of 3/2 can also be found in Cl doped I3 films (see Supporting Information, SI). It should be noted that in our case, the power indexes are discrete values. They are integral multiples of 1/2. Therefore, the values of power index can be well distinguished without ambiguity. The scanning electron microscope (SEM) images showing a 1×1 μm$^2$ film surface area are inserted in Fig. 2(a) and (b).

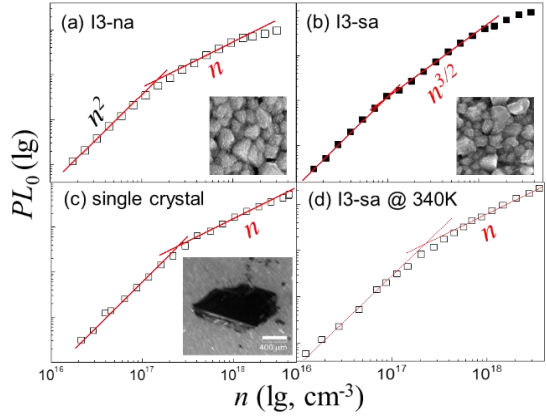

Figure 2. $PL_0$ versus excitation density $n$. (a) and (b) are the I3 films without/with annealing, respectively. The film thickness is ~250nm. The straight red lines are drawn with strict slop of 2, 3/2, and 1, for visualization. Corresponding SEM images are of 1×1 μm². The $PL_0$-$n$ curve for single crystal are also drawn, with density estimated according to a film of ~250nm thickness. (This estimation will not affect the power dependencies) The scale bar is 0.4 mm (c). The $PL_0$ of I3-sa film at 340K is recorded (d).

A specified photoproduct system can be identified by its density dependent behavior. The exciton and free carriers co-existing system has quadratic dependency at low excitation density. It become linear at high density. The critical density for power law conversion is decided by the exciton binding energy, $E_B$.[6] It can be well described by the Saha-Langmuir equation for a 3D semiconductor.[5] This is a simple but effective model for understanding the photoproducts inside semiconductors, since we consider the indexes only at lower and higher limits. The relative ratio of free carriers (electron or hole) in total excitation density, $x$, is written as,

$$\frac{x^2}{1-x} = \frac{1}{n}\left(\frac{2\pi\mu k_B T}{h^2}\right)^{3/2} e^{-\frac{E_B}{k_B T}} \qquad (1)$$

In this equation, the 1-$x$ is the corresponding ratio of exciton. $\mu$ is the reduced effective mass of ~0.15 $m_e$ (mass of electron [17]), $K_B$, $T$, and $h$ are the Boltzmann constant, the temperature, and the Planck's constant, respectively. The fluorescence includes both monomolecular emission of exciton and bimolecular recombination of free carriers. It can be written as

$$I(n) \propto A_1(1-x)n + A_2(xn)^2 \qquad (2)$$

In this expression, $A_1$ and $A_2$ are the decay rate of monomolecular and bimolecular emission terms, respectively. $x$ is from Eq. (1). By combination Eq.(1) and (2), the fluorescence quadratic increases at low excitation, which convert to linear dependency at high excitation density.[6] This means that the I3-na, single crystal, and I3-sa@340K have similar photoproduct system, which is System I.

The I3-sa presents a 3/2 power law dependency at high excitation density (Fig. 2 (b)). It can be attributed to the presence of a new emissive decay channel, ECC. We add a corresponding new term to Eq. (2),

$$I \propto A_1(1-x)n + A_2(xn)^2 + A_3(1-x)n \cdot xn \qquad (3)$$

$A_3$ is the decay rate of the ECC term. This term has a straightforward physical meaning that the ECC depends on both of the concentrations of free exciton and free carriers. This model generate the 3/2 power index at high excitation density. (A simulation can be found in SI.)

Next, the origination of System II, or ECC, must be analyzed. The following well-known factors are ruled out: (1) The Auger process appears at high excitation density over $10^{18}$ cm$^{-3}$,[6, 18-19] with different power index to $n$.[20-21] (2) The defects were estimated at the order of $10^{15}$ cm$^{-3}$,[4] which is even below the quadratic section of bimolecular recombination.[22] (3) From external appearance, the I3-na has no noticeable change on topological appearance to I3-sa, which means no external change can be assigned to the generation of ECC. (4) The unreacted PbI$_2$ domain, which is usually found in non-heat annealed films,[23] has no effect on photoproducts. This is because in the well-grown single crystals, the PbI$_2$ do not exist, to the best of our knowledge.[24] However, the I3-na film and the single crystal are similar in photoproducts. Therefore, by ruling out all other possibilities, a new intrinsic morphological origination must be included, which we assigned to the recently discussed intrinsic subgrain domain proposed by Frost et al.[8]

Besides the above analysis, the experimental proof also confirmed our assignment. The subgrain domains had been observed as twinning structure in tetragonal phase and disappear when the films are heated above the phase transition temperature.[15] The System I and II has similar dependence, showing that the ECC disappeared at cubic phase, and re-appeared after cooling down to room temperature. This is a solid base to our assignment since the photoproduct systems present same crystalline phase dependency. Besides, the results shows that the ECC generated in films when the newly made I3-na film was heat annealed to I3-sa. This is also the way to generate twinning structure in the inorganic perovskite films.[10] With both experimental proofs, we concluded that our assignment is reliable. The conclusion is summarized in Scheme 1.

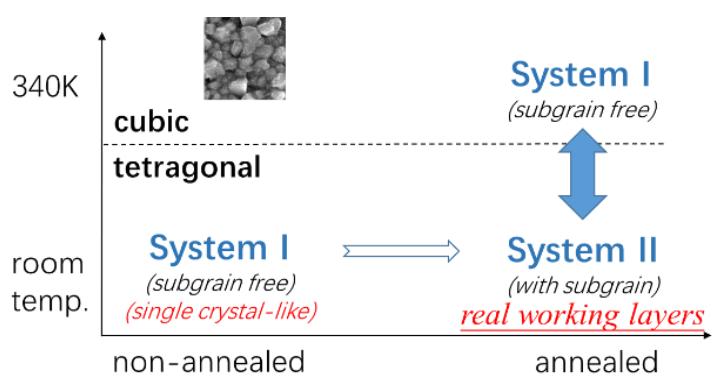

Scheme 1. The photoproduct systems for organolead perovskite films.

It should be noticed that the universality of subgrain domains are hardly to be proved. Besides the surface roughness which can hinder the observation, the domains may have variable size following square root law,[25-26] variable orientation, and irregular shape other than twinning domain.[13] However, these differences should not affect the general properties of efficient charge transportation, since the highly efficient organolead perovskite based solar cells can be easily repeated in numerous labs worldwide, regardless of these differences on morphology and processing methods. However, the current few studies, which clearly revealed subgrain morphology, require specialized sample films, such as flat-top surfaces.[12-13, 15] Our spectroscopic method provides an optical way to look into the films. It focus on the overall effect of subgrain structures and ignore above details. The method revealed the universality of ECC in heat-annealed films. Therefore, wide existence of subgrain domains in films for real working devices are then uncovered.

Frost et. al. suggested that the domain walls are the pathways for carrier transportation, so that the electron and hole are transported in different tunnels.[8, 27] It had been discussed that the subgrain domain boundaries may significantly affect the photophysical and electronic behaviors of organolead perovskite.[28] Such domain walls had been suggested with conductivity of 3 – 4 orders higher than the bulk material.[29] The study by Colsmann *et. al.* revealed charge extraction of photo-induced free carriers from the twinning ferroelectric structure.[13] Therefore, the long free carrier lifetime, slow recombination rate, and high conductivity in the perovskite films can be easily understood with the knowledge of such subgrain domains. Our discovery of System II uncovered the wide existence of subgrain domains. Such wide existence provides the morphological basis for understanding why the high carrier transportation property is general in organolead perovskite. We also suggest that our method and results could be applied for finding new perovskite materials with high performance.

**Conclusions**

In this report, we applied our newly developed density-resolved spectroscopic method and uncovered two distinct photoproduct systems in organolead perovskite. We found that when the films are newly made without heat annealing, the exciton and free carriers form a dynamically balanced photoproduct system, the System I. This is very similar to the single crystal. However, for the films with sufficient heat annealing processing, the ECC appears as a new decay channel. This forms the System II. Our analysis and experimental proof indicate that the ECC means the subgrain domain inside the crystal grains, which should be assigned to the ferroelastic /ferroelectric twinning structure. The results also mean the wide existence of such subgrain domains through the widely existed System II in films for real devices. This is a fundamental discovery for understanding the mechanism of high device performance, such as long carrier lifetime and diffusion length. We also suggest that such method and discoveries can be applied for searching new perovskite materials for high performance photovoltaic applications.

## ACKNOWLEDGMENT

This work supported by the National Basic Research Program of China 2013CB921904; National Natural Science Foundation of China under grant Nos. 61177020, 11134001, 11574009, 61575005, and 51402036.

Supporting Information

# Revealing two heat-annealing related photoproduct systems and widely existed subgrain domains in organolead perovskite

*Wei Wang,* [†,§] *Yu Li,* [†,§] *Xiangyuan Wang,* [‡] *Yang Liu, Yanping Lv,* [‡] *Shufeng Wang,\**[,†] *Kai Wang,* [‡] *Yantao Shi,\**[,‡] *Lixin Xiao,* [†] *Zhijian Chen,* [†] *and Qihuang Gong\**[,†] and Collaborative Innovation Center of Extreme Optics, Shanxi University, Taiyuan, Shanxi 030006, China.

[†]State Key Laboratory for Artificial Microstructure and Mesoscopic Physics, Department of Physics, Peking University, Beijing 100871, China.

[‡]State Key Laboratory of Fine Chemicals, School of Chemistry, Dalian University of Technology, Dalian, Liaoning 116024, China.

AUTHOR INFORMATION

**Corresponding Author**

* wangsf@pku.edu.cn; shiyantao@dlut.edu.cn; qhgong@pku.edu.cn

**Sample preparation:**

**I3-na and I3-sa:**
462 mg PbI2 (Sigma-Aldrich, 99.999%) was dissolved into 1 mL anhydrous dimethyl formamide (DMF, J&K, anhydrous, 99.8%) at 70°C and then were spin-coated on a quartz substrate at 5000 rpm for 10 s. After heating at 70°C for 30 min, they were cooled down to room temperature. Then the PbI2 films were dipped into a solution containing 10 mg $CH_3NH_3I$ in 1 ml 2-propanol (IPA, J&K, anhydrous, 99.5%) for 2 min. Then the films were rinsed with 2-propanol, and dried by spinning at 3000rpm for 30s. The I3-na was left as it was, while the I3-sa was further annealed at 70°C for 30 min.

**For I3Cl-sa:**
102.2 mg PbCl2 (Sigma Aldrich, 99.999%) and 174.9 mg $CH_3NH_3I$ (with a molar ratio of 1:3) into 0.5ml anhydrous DMF and then heated at 70 C for 1h. $CH_3NH_3PbI_{3-x}Cl_x$ films were fabricated by spin-coating of the precursor solution at 2000 rpm for 30 s. The I3Cl-sa was annealed at 90°C for 1 hr. and 100 °C for 25min.

**Crystallization of $CH_3NH_3PbI_3$:**
2.5 g PbAc2 and 10 ml hydroiodic acid (HI) 57 wt% in water were added into a tube. The tube was heated at 85 °C to dissolve the solution completely. Black precipitates came out when a solution, containing 2 ml HI and 0.597 g 40 wt% $CH_3NH_2$ in water (Merck), was added. Single crystals were grown along with the slow cooling of the solution from 85 °C down to 46 °C.

**More discussion for photoproduct conversion.**

*For exciton and free carriers dynamical co-existing photoproduct system, System I*

For perovskite system containing free carriers and exciton, such as freshly made I3-wa and single crystal, the dynamical co-existence of the two photoproducts follows Saha-Langmuir equation, shown in main text and Eq. S1 here,

$$\frac{x^2}{1-x} = \frac{1}{n}\left(\frac{2\pi\mu k_B T}{h^2}\right)^{3/2} e^{-\frac{E_B}{k_B T}} = \frac{1}{n} C(T, E_B) \tag{S1}$$

In this equation, $x$ is the ratio of free carriers in total excitation density, $n$. The corresponding $1-x$ is the ratio of exciton in total density, $n$. $\mu$ is the reduced effective mass of ~0.15 $m_e$ (mass of electron [1]), $K_B$, $T$, and $h$ are the Boltzmann constant, the temperature, and the Planck's constant, respectively. $E_B$ in this equation is the exciton binding energy.

From Eq. (1), we can derive the $x$ at low and high excitation limit. The Eq. (1) indicates that at the low limit where the free carriers are rich, $x \to 1$ and $x^2/(1-x) \to 1/(1-x) = C/n$. Then the $(1-x)$ is proportional to $n$. On the other hand, at high excitation density where exciton takes the main role, $x \to 0$ and $x^2/(1-x) \to x^2 = C/n$. Then the $x$ is proportional to $(1/n)^{1/2}$. This knowledge can also be found by theoretical calculation.[2] It should be motioned here that the C means the conversion density from the free carrier rich to the exciton rich situation.[3] The low and high excitation density limit means $n \ll C$ and $n \gg C$, respectively.

When the system containing photoproducts of exciton and free carriers, the fluorescence can be written as the Eq. 2 in main text, which is also copied here as S2

$$I(n) \propto A_1(1-x)n + A_2(xn)^2 \tag{S2}$$

In this expression, $A_1$ and $A_2$ are the decay rate of mono-molecular and bi-molecular emission terms, respectively. $x$ and $n$ are taken from Eq.(S1). At low and high excitation density, the Eq.(S2) include two components representing the emissive decay of exciton and bimolecular recombination, respectively. At low and high excitation density, $x \to 1$ and 0, where the S2 will be S3 and S4, respectively.

$$I(n) \propto \frac{A_1}{C} \cdot n^2 + A_2 n^2 = \frac{1}{C}(A_1 + A_2 C)n^2 \tag{S3}$$

$$I(n) \propto A_1 n + A_2 C n = (A_1 + A_2 C)n \tag{S3}$$

The equation clearly shows quadratic dependency at low excitation density and linear dependency at high density. With this knowledge, an exciton and free carrier balanced system can be derived.

### *With the presence of exciton-carrier collision (ECC), System II*

When ECC appears, one more term should be added to Eq. S2

$$I \propto A_1(1-x)n + A_2(xn)^2 + A_3(1-x)n \cdot xn \qquad (S4)$$

As the former analysis, at high excitation density, the free carrier density, $xn \propto n^{1/2}$. On the other hand, the exciton linearly increases to the density, $(1-x)n \propto n$. Then the ECC term will present a 3/2 power law dependency to the density. At high excitation density, it becomes the major source of density dependent fluorescence. At low excitation density, since the third term is small comparing to the first two terms, the $PL_0$ still have the quadratic dependency to the density. A simulation can be seen in the following part.

*Simulation of density dependent photoproduct behavior*

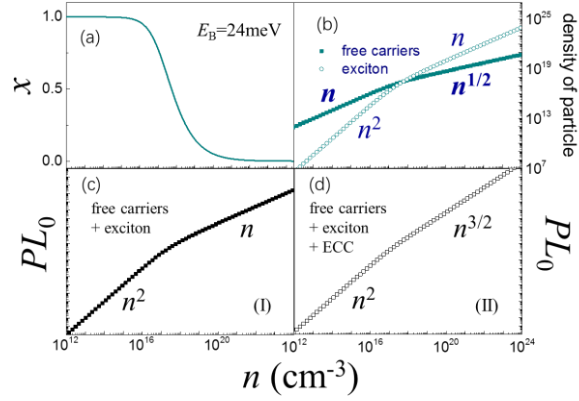

**Figure S1**. The simulation of density-dependent photoproducts inside perovskite films. (a) The relative ratio, $x$, of free carriers in total excitation density, according to Saha-Langmuir equation. (b) The simulated density of free carriers and excitons. (c) The simulated density-resolved $PL_0$ when the free carriers and exciton dynamically co-exist inside perovskite, and (d) when ECC become dominant in the system.

However it should be noticed that, from low density to high density, the transition from quadratic to 3/2 power law dependency in Fig. S1(d) may not be a smooth turning. As simulated below, we present three curves showing that $A_3=A_2$, $A_3=10A_2$, and $A_3=0.1A_2$. The results are shown in Figure S2. Such phenomena can be slightly observed in Fig. 2(b) and Fig. S3.

The power index of the photoproducts and $PL_0$ towards $n$ are summarized in Table S1.

**Table S1**. The density-dependent photoproducts and $PL_0$.

| density | fc | exci | ECC | $PL_0$ |
|---|---|---|---|---|
| $n < 10^{17} \text{cm}^{-3}$ | $\sim \boldsymbol{n}$ | $\sim \boldsymbol{n^2}$ | $\sim n^3$ | $\sim n^2$ |
| $n > 10^{17} \text{cm}^{-3}$ | $\sim n^{1/2}$ | $\sim n$ | $\sim \boldsymbol{n^{3/2}}$ | $\sim n^{3/2}$ |

fc: free carriers. exci: exciton. The dependency of exciton and free carriers can also be found in Ref.[2]. The main photoproducts and decay channels at each density range are marked with bold font.

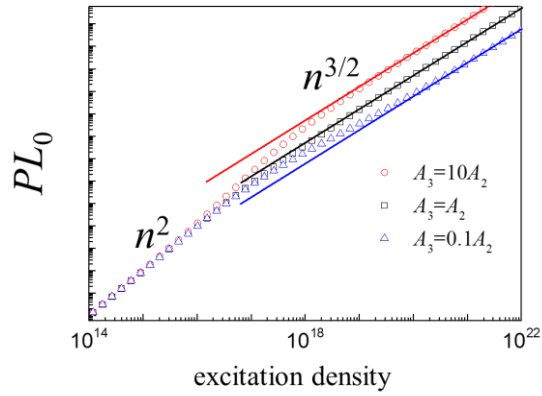

**Figure S2**. The simulation of Eq. S4, supposing $A_1$=0. The $A_2$ and $A_3$ are of different ratio.

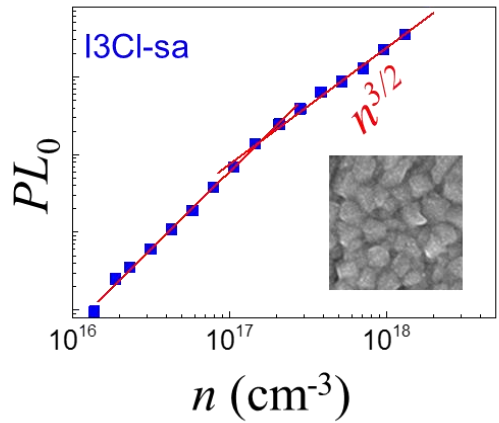

**Figure S3.** The $PL_0$ vs $n$ for chlorine doped perovskite films. The SEM image of 1×1 μm$^2$ are inserted in the figure.